\documentclass[prl,aps, draft]{revtex4}
\begin{document}
\large
\begin{titlepage}
\title{Ultrahigh Energy Neutrals From Extreme Magnetic Flares}
\author{David Eichler}
\affiliation{Physics Department, Ben-Gurion University,
Beer-Sheva}

\begin{abstract}

 It is shown that bulk acceleration
during reconnection of extremely strong magnetic fields near
compact objects can accelerate ions to Lorentz factors of $\sim
10^2 \sigma^{3/5}$ under general conditions, where $\sigma$, the
magnetic energy per current-carrying proton rest energy, can
approach $10^{15}$. For magnetar-type fields, neutrons and
neutrinos can be generated at potentially detectable levels via
hadron polarization. Ultrahigh energy photons can also be emitted
and escorted from the high field region by Poynting flux.

\end{abstract}
\maketitle
\end{titlepage}

\section{Introduction}

Magnetic flares on the surfaces of  strongly magnetized neutron
stars (magnetars) are believed to be responsible for repeating
soft gamma ray bursts (SGR events)  \cite{Wood01}.
While such events are detected primarily in soft $\gamma$-rays,
there could be much more energetic quanta produced as well, and
there could be more classes of events than those that observations
have heretofore identified. Perhaps such events cause other
sub-classes of gamma ray bursts as well. As is the case for solar
flares and related transient coronal heating phenomena, some
magnetar outbursts might be dominated by emission from
magnetically trapped plasma, while others might direct most of
their energy outward in the form of Poynting flux, and could be
much quieter in low energy $\gamma$-rays.  Moreover, while some
flares may occur on the surfaces of established magnetars, others
might occur when a strongly magnetized compact object (SMCO)
collapses to a black hole, which, by the no-hair theorem, must
swallow or otherwise rid itself of its large scale magnetic field.

This letter suggests that ultrahigh-energy quanta such as cosmic
rays, neutrons, neutrinos and photons could all be created in such
events at detectable levels, and might even be a common feature to
the various types of imaginable events. The primary acceleration
mechanism analyzed here, particularly appropriate to SMCO
magnetospheres,  is
 bulk acceleration of plasma during magnetic reconnection
in the magnetospheres of extremely magnetized compact objects such
as SGR's. It parallels solar flares and magnetospheric substorms
in the Earth's magnetotail, but transplants the mechanism to SGR's
 and other SMCO's, whose huge magnetic
fields, $B\gg 10^{14}$G, combined with huge gravitational fields,
create favorable conditions for attaining ultrahigh energies. The
mechanism faces no injection problem, because all the particles
are accelerated to ultrahigh energies. Pair creation need not
short out the electric field, which is perpendicular to {\bf B},
and the pairs do not significantly affect the maximum attainable
ion energy as long as their mass density in the zero electric
field frame is less than that of the baryons.  The key is the
rarefaction of the magnetospheric plasma, which is accomplished by
the strong gravitational field of the compact object (somewhat
analogous to the rarefaction in the Earth's magnetotail).

A key feature of this acceleration mechanism is its ability to
escort high energy quanta: Because the bulk Lorentz factors are so
huge, the magnetic field in the zero electric field  frame is
greatly reduced relative to the static magnetospheric value.
Moreover, the energy of emerging high energy quanta is much
smaller in the fluid frame than in the lab frame. Thus, high
energy photons that would pair produce (and high energy charged
particles which would synchrotron cool or curvature radiate) in
the strong static magnetic field of the SMCO can be escorted
through it by a high Lorentz factor flow.

Consider a strongly magnetized compact object  (SMCO,  e.g.
neutron star, collapsar, steadily accreting black hole) of radius
R  with a field strength B expressed as B=$\epsilon B^*$ where
$B^*$, the maximum field strength imaginable for an object of mass
M and associated Schwarzschild radius $R_s = 2GM/c^2$, is given by
$ {B^*}^2R_s^3/8\pi = Mc^2 $ Because the gravitational field
allows a negligible thermal scale height, the plasma density well
above the surface is determined purely by electrodynamics, and is
of order $B/4\pi R\beta$, where $\beta c$ is the velocity of the
charge carriers. The characteristic current associated with the
field is of order $c B/4\pi R$, and the minimum proton or electron
number density associated with this charge density, $n=j/e\beta
c$, is $n   \sim B/4\pi R e\beta c =  2 \times
10^{17}B_{15}R_6/\beta\, \rm cm ^{-3} $ where numerical subscripts
of any quantity refer to powers of ten by which the quantity is to
be multiplied when expressed in cgs units. If the current is
carried by protons, then  the energy per current carrying proton,
$B^2/8\pi n$, is of the order of
\begin{equation}
\sigma m_pc^2 \equiv eBR = (4\pi)^{1/2}\epsilon \beta
\left(e^2/Gm_p^2\right)^{1/2}(R/R_s)m_pc^2 \equiv \epsilon \beta
 (R/R_s)\sigma^* m_pc^2
\end{equation}
whence
\begin{equation} \sigma = 5 \times 10^{18} \epsilon \beta
(R/R_s).
\end{equation}

While the quantity $\epsilon$ is sure to be small, it can be as
high as $10^{-5}$ for magnetars, and the quantity $\epsilon
(R/R_s)$ as high as $0.3 \times 10^{-4}$.  The energy per current
carrying proton can be as high as  $10^{14.5}m_pc^2$. If a
"failed" or spinning down magnetar collapses to a black hole, then
just prior to completing this collapse, it would have a field of
about an order of magnitude higher, and  $\epsilon (R/R_s)\sim
10^{-4}; \sigma \sim 10^{15}$.  In practice, a single species
plasma would result in extremely high electric fields and the
plasma is likely to be quasi-neutral. In this case,  the
multiplicity $\xi$, i.e. the ratio of the actual plasma mass
density to the minimum needed to provide the curl of {\bf B}, is
likely to be m greater than unity. In pulsars, the charge
multiplicity is unknown, but could be as high as $10^4$  or more
\cite{Hi01}, and  if, as will be assumed here, the current
carriers include ions,  the mass multiplicity is even less
certain.

If a significant fraction of the magnetic energy density  is
stored in a solenoidal component of this field, it could be
released by reconnection and current dissipation.

The total magnetic energy of a magnetar can approach $10^{47}$
ergs. A SMCO that collapsed through the magnetar stage could
undergo a further factor of 30 increase in total field energy. If
such objects formed every $10^3$ years or so per galaxy, they
could produce up to about $10^{44}$ergs/yr-Mpc$^3$ in ultrahigh
energy cosmic rays, a considerable fraction of the total. This is
motivation to consider the maximum energy in greater detail, as
done below.

\section{Bulk Acceleration}

A sudden eruption resulting from magnetic energy release in a high
$\sigma$ magnetic configuration probably results in bulk motion
with a Lorentz factor $\Gamma$ of order the magnetosonic Lorentz
factor $(\frac \sigma \xi)^{1/2}$. It may be a slow mode shock
propagating away from a reconnection point \cite{Bla94,Bla96}, or
a sudden decompression of twisted magnetic field lines.  The zero
electric field  (ZEF)  in this picture is established by a
"piston" of plasma ejected from a magnetic reconnection site.
Material swept up just ahead of the piston has about the same
Lorentz factor to the piston itself.
 Bulk acceleration of
plasma by a relativistic outflow has been considered by Michel
(1984) \cite{Mic84}, who finds that the maximum energy attained by
a given particle is of order the initial energy times
$\sigma^{2/3}$. Here, since the geometry is likely to be messy, we
adopt a simplified  but general approach, and generalize Michel's
basic conclusion, more or less, to other geometries.

To simplify geometric considerations, consider a uniform magnetic
field {\bf B}=$10^{15}B_{15}${\bf b}, where {\bf b} is the unit
vector in the field direction,  and a frame that moves
perpendicular to {\bf B} with Lorentz factor $\Gamma$ relative to
the frame of the compact object, hereafter called the lab frame.
In the  ZEF  frame, the field is $B'= B/\Gamma$ and the Lorentz
factor of the particle is denoted by  $\gamma'$.

{\it Adiabaticity:} The proton gyrofrequency  in the ZEF frame is
\begin{equation}
\omega'_g = eB'/\gamma'mc = 1\times 10^{19}B'_{15}s^{-1}/\gamma'.
\end{equation}
The characteristic  proper acceleration time $\tau'_{acc}$ of the
ZEF  frame to a Lorentz factor of $\Gamma$, from a comparable but
lower value, is at most the hydrodynamic proper timescale
$R/\Gamma c$, i.e.
\begin{equation}
\tau'_{acc} \le   R/c\Gamma. 
\end{equation}
The condition that protons drift with the ZEF  is $\omega'_g
\tau'_{acc}\gg 1$, i.e.

\begin{equation}
\Gamma^2\gamma' \le \sigma.
\label{7}
 \end{equation}
 So for  $\Gamma \gamma'^{1/2}\le
(10^{14.5}B_{15}R_6/c)^{1/2}$,
 the protons should respond
adiabatically 
to the electromagnetic impulse, and drift in the zero electric
field frame.  Thus, if large amplitude magnetosonic motion is
generated at the Lorentz factor $({\frac\sigma\xi})^{1/2}$, the
individual Lorentz factors of the typical plasma ions may be
comparable. (Note that if a particle at rest in the lab frame is
suddenly picked up, then $\gamma'=\Gamma$, and the maximum
attainable lab frame energy is $\sigma^{2/3}$, similar to the
result obtained by Michel \cite{Mic84} for the specific case of
stationary radial winds.)

Particles that begin with  large Lorentz factors  $\gamma_i$ in
the lab frame, $\gamma_i \gg 1$, can ultimately achieve energies
even greater than $\sigma^{2/3}mc^2$ if $\Gamma$ remains below
$\sigma^{1/3}$. Ultra-relativistic MHD turbulence where
independent cells attained Lorentz factors of $\Gamma$ could
accelerate ions via multiple encounters to
\begin{equation}
\gamma_{max}= 2\Gamma \gamma'_{max} = 2\sigma /\Gamma.
\end{equation}
 Though analogous to
second order Fermi acceleration, such a process would be nearly as
efficient as a first order process because the changes in energy
are in large increments.

The synchrotron loss time for a charged particle of mass m is
given by
\begin{equation}
{\tau'_{syn}}^{-1} = \gamma' e^4 B'^2/3{m}^{3}c^{5}=\gamma'
e^4B^2/3m^{3}c^5\Gamma^2.
\end{equation}
The condition that  $\tau'_{syn}\ge R/\Gamma c$ can be written as
\begin{equation}
\gamma'\le \gamma'_{cool}\equiv 3\Gamma^3c/\sigma r_{o}
\omega_g\le 1 \label{10}
\end{equation}
where $r_{o}$ is the classical electromagnetic radius of the
particle. Writing the maximum lab frame energy  as $2\gamma'
\Gamma = 2(\gamma'/\Gamma^3)^{1/5}(\Gamma^2\gamma')^{4/5}$ and
using equations (\ref{7}) and (\ref{10}),  one concludes that
synchrotron losses are small provided that
\begin{equation}
2\gamma' \Gamma \le 70 \sigma^{3/5}(m/m_p)^{2/5}B_{15}^{-1/5}
\label{11}
\end{equation}

We conclude that near SMCO's, protons can attain  Lorentz factors
of nearly  $\sim 10^{11}$  by this mechanism.

\section {Meson Production}

{\it Polarization Induced Mesons:} If a hadron propagates with a
Lorentz factor $\gamma'$ relative to the zero electric field
frame, then there is an electric field of $\gamma' B'$ in the rest
frame of the hadron. When this field exceeds $\frac35 T/e$
[$\frac{3}{4}T/e$], where T is the QCD string tension $ 1 \times
m_p^2c^3/h $ ($0.16 GeV^2 $ in units where $ \hbar=c=1$), a proton
[neutron] is electrically polarized enough to overcome QCD
confinement and the hadron is stretched into a long flux tube over
a timescale of $10^{-23}$s. Virtual quark pairs will materialize
along this color flux tube, and probably the negative and positive
quarks are each pulled to opposite sides.  We therefore conjecture
that the resulting pions are mostly charged and ultimately decay
into neutrinos. The mechanism operates even at constant velocity,
and  even on neutrons, but is otherwise reminiscent of  curvature
pion radiation \cite{Be95}.

 This threshold occurs at a Lorentz factor
(still in the zero electric field frame) of
\begin{equation}
\gamma_{th}'\sim {m_p}^2c^3/eh{B'}_{\bot} = 10^{4.5 }{B'_{\bot
15}}^{-1}.
\label{12}
\end{equation}
Any hadron  that is injected at $\gamma'\ge \gamma_{th}$ generates
additional mesons.
 Because the proper time is much less
than the decay time of a pion,  pions reproduce themselves as long
as their Lorentz factor in the zero electric field frame exceeds
$\gamma_{th}'$.

A hadron could  establish sufficient cross-field motion to exceed
the threshold $\gamma'_{th}$ established by  equation (\ref{12})
via inertial forces while following curved field lines. Also, a
UHE neutron, once produced (by photopion production, say), could
coast into some other cell of ultrarelativistic turbulence where
its local Lorentz factor could exceed  $\gamma'_{th}$.

 For simplicity,  consider an ion that
has already been accelerated to UHE energies that is now exiting
the acceleration region along static curved field lines, so that
that the ZEF frame is (only in the present example) the lab frame.
Ions that move with Lorentz factor $\gamma_{\|}$ along curved
magnetic field lines with a radius of curvature $R_c$ experience a
perpendicular acceleration of $a_{\bot} = c^2/R_c$. In the frame
of the ion, where the perpendicular acceleration $a'_{\bot}$ is
$a'_ {\bot} = \gamma_{\|}^2 a_{\bot}$,
the force is $F' = {\gamma_{\|}}^2 m_i c^2/R_c$. The differential
electromagnetic force on a proton is 5/3 of the total force, and
when this exceeds the tension in the color flux tubes that bind
oppositely charged quarks, the hadron emits a steady steam of
mesons until it has slowed down to below the threshold for this
process.  Thus for
\begin{equation}
\gamma_{\|} \ge \gamma_{c,th}\equiv  (R/r_p)^{1/2} =2 \times
10^{9}R_{c6}^{1/2},
\end{equation}
where $r_p$ is the radius of the proton, one Fermi, pions would be
emitted, so the Lorentz factor of parallel motion is limited to
this in the zero electric field frame.  By equation (7) above,
this threshold Lorentz factor  for polarization-induced pion
production can be achieved for $R_{c6}B_{15} \sim 1$. For protons,
however, electromagnetic  curvature radiation would limit  the
Lorentz factor to $(R_cm_pc^2/ e^2)^{1/3}$, which is  below the
above limit.

  It is easier, on the other hand,
for neutrons to establish enough cross-field motion to meet the
threshold condition (\ref{12}). Even a $\pi^0$, moving at a
$\gamma$ of $10^{10}\gamma_{10}$, travels $\sim 2 \times
10^4\gamma_{10}$ cm before decaying.  Over this distance, the
field would curve by $2 \times 10^{-2}R_6$ radians, and this is
enough to cause the $\pi^0$ to further cascade. Thus, neutral
hadrons with $\gamma\gg \gamma_{c,th}$  loosely follow field
lines. It can be shown that they lose about $\delta\theta$ of
their original energy to meson production as the field curves
through an angle $\delta\theta$. Neutrons with $\gamma\ge
\gamma_{c,th}$  could thus escape the system along moderately
polar field lines without losing most of their energy to mesons,
though they might lose of order half or so.

{\it Photopions and Neutrons:} Any neutron stars  that emit X-rays
at $\eta L_{ed}$ where $L_{ed}$ is  the Eddington luminosity  can
convert protons to neutrons and accompanying $\pi^+$'s within
about $\eta 10^{8.5} cm$ \cite{Eichl78}. Magnetars, which emit
steadily at about $10^{-3}L_{edd}$, can thus convert over
$10^{-1}$ of UHE protons above $10^{14.5}$eV  produced near their
surface. The fraction can be higher if the protons undergo many
ultrarelativistic  oscillations while trapped on a field line, and
when the X-ray luminosity goes up during the flare.

 Magnetars also emit optical and near IR
radiation \cite{Ke02,Is02,Ka02}. If due to coherent plasma
instabilities in the magnetospheric currents \cite{Eic02}, this
should be a generic feature of magnetars, although in most cases
such emission would be obscured by interstellar dust. Corrected
for reddening, the intrinsic optical luminosity of the magnetar
4U0412 is probably about $10^{33}$ erg/s. These photons, energy of
order 1 to 2 eV, may be generated by coherent processes near the
magnetar surface, and their density is of order
\begin{equation}
n_{\gamma} = 10^{21}cm^{-3}L_{33}R_6^{-2}.
\end{equation}

The cross section for photopion production above the  threshold
$300 MeV$ in the frame of the proton is $3 \times 10^{-28}cm^2$.
The optical depth to photopionization by optical photons over a
length of $10^6 R_6 $cm is thus above $10^{-1}$ per passage, and
neutrons can be formed in significant quantities by optical
photons from  protons with lab frame Lorentz factors of order
$10^{8.5}$. Iron nuclei would be photodisintegrated  at Lorentz
factors about $10^4$ ($10^7$) by X-ray photons (optical photons),
and this is another way to produce free neutrons at $\Gamma \ge
10^4$ ($\Gamma \ge 10^7$).

\section{Photon Pickup}
  A photon with lab frame energy $E_{ph}$ has an energy in the
frame of the piston of $\Gamma E_{ph}$  to within a geometric
factor. Photons with lab frame energy above $ (B_{QED}/B)m_ec^2$
pair produce  in the ZEF frame. The pairs   emit  synchrotron
radiation at a frequency of $2\times 10^7 \gamma'^2B_0'$ where
$B_0'$ is the ZEF field strength in Gauss. As they slow down, they
emit an increasing number of photons per decrease in ln $\gamma'$.
Assuming a classical approximation
\begin{equation}
d\gamma'/dt'= \gamma'^2\sigma_T B'^2/8\pi m_ec
\end{equation}
\begin{equation}
d\theta/dt = eB'/\gamma'm_ec, \end{equation}
 it is straightforward to show that the photons emitted
after 1/4 of a gyroperiod ($\theta=\pi/2)$, i.e. perpendicular to
the fluid velocity) are emitted in the ZEF frame at an energy
$\epsilon'$ of
\begin{equation}
\epsilon' = \hbar \gamma'^2eB'/mc = 3m_ec^2/\alpha \pi.
\label(20)
\end{equation}
 The energy is
the lab frame is thus
\begin{equation}
 \epsilon = 67 \Gamma MeV
 \label{21}
\end{equation}
which, for the magnetosonic value of $\Gamma$,  $\Gamma =
(\sigma/\xi)^{1/2}$, can exceed $10^{15}(\sigma_{15}/\xi)^{1/2}$
eV. Photons this energetic could not escape from static SMCO
magnetospheres. However, they survive in the high $\Gamma$
outflow, which has a much lower field than the lab frame. If the
ZEF frame  is outward going, the Poynting flux escorts the high
energy photon out to large distances. A photon with the energy
$\epsilon'$ given by equation (\ref{20}) is below the pair
production threshold as long as $B'\le \pi \alpha B_ {QED}$. The
final lab frame energy depends only on the Lorentz factor of the
fluid element that finally escorts it to a region where the static
field is too weak for pair production.

 The number of second generation
photons emitted per first generation photon is of order
$B/B_{QED}$. This could be somewhat less than unity for solar mass
SMCO's, suggesting that a quantum calculation is called for.

\section{Discussion}

An actual flare from a magnetar or other SMCO may be quite
complicated,  because the photon flux increases by many orders of
magnitude during the event,  because the Lorentz factor of the
ultrarelativistic turbulence is likely to vary widely, and because
the flare itself would distort the preexisting magnetosphere in
ways that are hard to predict. As such, the preceding discussion
has been deliberately general.  Using the considerations of
previous sections, we can suggest a sample scenario using
parameters at the beginning of the flare keeping in mind the
uncertainties: Magnetic reconnection leads to a large scale motion
of field lines. The Lorentz factor $\Gamma_1$ is as much as
$(\sigma/\xi)^{1/2}= 10^{7.5}(\sigma_{15}/\xi)^{1/2}$. This is
enough that protons dragged by the field lines are photo-converted
to neutrons which then pass freely to regions where the ZEF
magnetic field is of order $10^{15}/\Gamma_2$G. There they emit
mesons until decelerating to the threshold
$\gamma'_{th}=10^{4.5}B_{15}^{-1}\Gamma_2$ defined by equation
(\ref{12}). In doing so, they can convert back to protons. If
$\gamma'_{th}\le\sigma/\Gamma^2$, as per condition (\ref{7}), they
may
 reverse direction in the ZEF frame
and cool down to $\gamma'_{cool}$.  The final  lab frame Lorentz
factor after this reversal is then given by the right hand side of
equation (\ref{11}), and, if the energetic protons eventually
attain isotropy in the fluid frame, they have a flat energy
distribution out to this maximum value in the lab frame. (If, on
the other hand, the reconstituted proton does not satisfy equation
(\ref{7}), it passes through that cell of turbulence or
synchrotron breaks until it has slowed enough in the second fluid
frame to do so, or else escapes the system.)

Alternatively, though for a more specialized geometry, we could
suppose that protons pass through a slow shock \cite{Bla94}, the
Petschek model of magnetic field line reconnection, in which
magnetic energy is converted to particle kinetic energy.
Downstream of the shock, the fluid moves in the plane of the shock
with bulk Lorentz factor $\Gamma$. Although we have no rigorous
theory of collisionless slow  mode shock structure (much less in
the ultrarelativistic limit) we assume that the ions are picked up
suddenly (i.e. $\gamma'=\Gamma$) subject to the constraint of
equation (\ref{7}) and conjecture that $\Gamma \sim \sigma^{1/3}$
and that the particles  attain lab frame energies of up to
$\sigma^{2/3}$ or the right hand side of equation (\ref{11}) if it
is less. Note that equation (\ref{7}) expresses a necessary
condition for the validity of MHD, whereas reference \cite{Bla94}
assumes MHD. Thus, the former can be more restrictive in high
$\sigma$ environments. It expresses the restriction that the shock
thickness cannot exceed the dimensions of the region.

Either of the above variations allows final Lorentz factors in the
lab frame well above $10^9$, and the proton could reconvert to a
neutron upon exiting the region. There is thus some chance that
neutrons so produced could arrive intact at Earth from a distance
of 10 Kpc.

Given a number density of current carrying protons of
$10^{18}B_{15}/R_6$ over typical dimensions of $10^6$cm, the
number of available protons could be of order $10^{35}$ or more.
(Note that if the burst lasts more than a dynamical timescale, the
current carriers turn over, and the time integrated total can be
more than at a given instant.) If a significant fraction of them
are neutronized at Lorentz factors $\Gamma$ above $10^{9}$, then
they propagate 10 Kpc before decaying. At a distance of 10 Kpc, a
burst of $10^{44}$ erg that puts $10^{-4}$ of its energy into
$10^{34}$ neutrons at $\Gamma = 10^9$ creates a fluence of
$10^{-2}/km^2$. A collecting area exceeding 100 $km^2$ might have
had some chance of detecting UHE neutrons from  the Aug. 27, 1998
flare. Similarly, a $10^3km^2$ array might have had the chance
from the April 18, 2001 flare, which emitted about $10^{43}$ ergs.

The charged pions emitted typically cool until they marginally
satisfy equation (\ref{10}) suitably adapted to pions, i.e. until
$\gamma'=3 \Gamma^3c/r_{\pi,o}\sigma\omega_{\pi,g}\sim1.3\times
10^7\Gamma^3/\sigma$, and they can produce extremely energetic
neutrinos, of order $E_{\nu}=\gamma' m_{\pi}c^2/4 \sim (5\times
10^{14}\Gamma^3/\sigma)$eV.
 The neutrino fluence at Earth that would have resulted from the Aug. 27
flare  would have been $10^{-2}$ erg/cm$^2$ if  all the observed
flare energy had gone into neutrinos. Several percent of this in
the range 1 TeV $\le E_{\nu}\le 10^3$ TeV is
 detectable with 1 $km^3$ detectors.

A photon fluence of $10^{-2}$ erg/cm$^2$, the energy equivalent of
the soft $\gamma$-rays from the Aug. 27 giant flare from 1900+14,
would produce a huge signal in MILAGRO if the photon energies
exceed 1 TeV. If energies approach that given by equation
(\ref{21}) with $\xi=1$, then the maximum energy can exceed
$10^{15}$ eV and this should be accessible with  air shower
detection.

As the flare from a magnetar progresses, and the photon luminosity
increases, the emission  from a vibrating magnetosphere should be
increasingly dominated by the emission of pairs from swept-up
photons. However, this is not necessarily true if the energy is
released in outward Poynting flux or if a black hole forms during
the collapse of a SMCO; in these cases there may be little
thermalization of UHE emission.

Altogether, ultrahigh energy emission could be a major component
of the radiative output of giant flares on SMCO's. If it could
carry {\it most} of this output, it could conceivably allow
magnetospheric rearrangement  that is quiet in soft and medium
energy $\gamma$-rays,  which is suggested by the sudden changes in
the spin down rate observed for some magnetars.  However, this
possibility requires further investigation.

I gratefully acknowledge very  helpful discussions with D. Faiman,
M. Gedalin, E. Gedalin, A. Mualem, A. Levinson,  Y. Lyubarsky, B.
Svetitsky, C. Thompson and V. Usov. This research was supported by
an Adler Fellowship administered by the Israel Science Foundation,
and by the Arnow Chair of Astrophysics.

\end{document}